\documentclass[aps,prc,twocolumn,superscriptaddress,showpacs,amsmath,floatfix,
nofootinbib]{revtex4}
\usepackage{graphicx}
\usepackage{bm}

\begin{document}


\title{Structure of upper $g_{9/2}$-shell nuclei and shape effect
in the $^{94}$Ag isomeric states}

\author{K.~Kaneko}
\email{kaneko@ip.kyusan-u.ac.jp} \affiliation{Department of Physics,
Kyushu Sangyo University, Fukuoka 813-8503, Japan}
\author{Y. Sun}
\email{ysun@nd.edu, sunyang@sjtu.edu.cn} \affiliation{Department of
Physics, Shanghai Jiao Tong University, Shanghai 200240, P. R.
China} \affiliation{Joint Institute for Nuclear Astrophysics,
University of Notre Dame, Notre Dame, Indiana 46556, USA}
\author{M. Hasegawa}
\affiliation{Institute of Modern Physics, Chinese Academy of
Science, Lanzhou 730000, P. R. China}
\author{T. Mizusaki}
\affiliation{Institute of Natural Sciences, Senshu University, Tokyo
101-8425, Japan} \affiliation{Institut de Physique Nucl\'{e}aire,
Universit\'{e} Paris-Sud F-91406 Orsay CEDEX, France}

\begin{abstract}

Using a shell model which is capable of describing the spectra of
upper $g_{9/2}$-shell nuclei close to the $N=Z$ line, we study the
structure of two isomeric states $7^{+}$ and $21^{+}$ in the odd-odd
$N=Z$ nucleus $^{94}$Ag. It is found that both isomeric states
exhibit a large collectivity. The $7^{+}$ state is oblately
deformed, and is suggested to be a shape isomer in nature. The
$21^{+}$ state becomes isomeric because of level inversion of the
$19^{+}$ and $21^{+}$ states due to core excitations across the
$N=Z=50$ shell gap. Calculation of spectroscopic quadrupole moment
indicates clearly an enhancement in these states due to the core
excitations. However, the present shell model calculation that
produces the $19^{+}$-$21^{+}$ level inversion cannot accept the
large-deformation picture of Mukha {\it et al.} in Nature {\bf 439},
298 (2006).

\end{abstract}


\maketitle

\section{Introduction}\label{sec1}

The structure study of $N\approx Z$ nuclei is one of the current
topics in nuclear physics. For the upper $g_{9/2}$-shell $N\approx
Z$ nuclei, perhaps the most interesting aspect is the occurrence of
high-spin isomers. For decades, spin-gap isomers have been predicted
by shell model calculations for nuclei close to the double-magic
$^{100}$Sn \cite{Ogawa}. Spin-gap isomers have recently been
observed in some heavy $N\approx Z$ nuclei, for example, in
$^{95}$Ag, $^{95}$Pd, and $^{94}$Pd \cite{Marginean}. From the shell
model point of view, it is understood that these isomers are formed
by an extra binding energy due to large attractive proton-neutron
({\it pn}) interaction in the maximally aligned particle-particle or
hole-hole configurations. Thus, one sensitive test for effective
interactions in the shell model is a quantitative description of
these high-spin states and their decay path.

The study of $N\approx Z$ nuclei has important implications in
nuclear astrophysics. It has been suggested that in x-ray binaries,
nuclei are synthesized via the rapid proton capture process (rp
process) \cite{Wormer,Schatz1}, a sequence of proton captures and
$\beta$ decays responsible for the burning of hydrogen into heavier
elements. The rp process proceeds through the exotic mass region
with $N\approx Z$. New reaction network calculations \cite{Schatz2}
have suggested that the rp process can extend up to the heavy Sn-Te
mass region, involving the nuclei that we study in the present
paper. Since the detailed reaction rates depend sensitively on the
nuclear structure, information on energy levels of relevant nuclei
is thus very useful. As emphasized by Schatz {\it et al.}
\cite{Schatz1}, understanding the so-called waiting point nuclei is
particularly important. Furthermore, if isomeric states exist in the
nuclei along the rp process path, the astrophysical significance
\cite{Sun05,Sun} could be that the proton-capture on long-lived
isomers may increase the reaction flow, thus reducing the timescale
for the rp process nucleosynthesis during the cooling phase.

Because of the recent experimental successes, the high-spin
$I^\pi=21^{+}$ and low-spin 7$^{+}$ isomers in the odd-odd $N=Z$
nucleus $^{94}$Ag have become a discussion focus
\cite{Commara,Plettner,Mukha041,Mukha042}. In this nucleus, the
high-spin 21$^{+}$ isomer has a high excitation energy of 6.7(5) MeV
with a notably long half-life of 0.39(4) s, and is open to $\beta$,
one-proton, and two-proton decays \cite{Mukha05,Mukha06}. Although
the shell model calculations with the empirical effective
interaction in the restricted (1$p_{1/2}$, $0g_{9/2}$) model space
could reproduce the energy levels and high-spin isomers in
$^{95}$Ag, $^{95}$Pd, and $^{94}$Pd, it failed to predict the
isomerism of 21$^{+}$ state in $^{94}$Ag \cite{Mukha041,Plettner}.
On the other hand, it has been shown that the large-scale shell
model calculations with the extended model space ($0g_{9/2}$,
$1d_{5/2}$, $0g_{7/2}$, $1d_{3/2}$, $2s_{1/2}$) can obtain a
21$^{+}$-19$^{+}$ level inversion, which suggests that the core
excitations across the $^{100}$Sn shell-closure play a crucial role
in generating the 21$^{+}$ isomer with such a long half-life
\cite{Plettner}.

The analysis on one-proton decay \cite{Mukha05} and two-proton decay
\cite{Mukha06} data of the 21$^{+}$ isomer in $^{94}$Ag has
suggested a strong deformation picture for this state. The authors
of Ref. \cite{Mukha06} claimed that the unexpectedly large
probability for the proton radioactivity could be attributed to a
large deformation of the parent nucleus with a prolate shape.
However, it is questionable that the 21$^{+}$ isomer in $^{94}$Ag
is strongly deformed because the ground states of nuclei in this 
region are very weakly deformed. 
It has recently been pointed out \cite{Roeckl} that the
large-deformation claim remains a puzzle. On the other hand, the
half-life of the low-spin 7$^{+}$ isomer has been measured to be
0.59(2) s \cite{Plettner}, while the excitation energy is yet to be
determined.

However, the question why these states become isomeric and what the
nature of the isomerism is, has not been thoroughly addressed.
Recently, we have investigated the structure of low-spin isomeric
states in the odd-odd $N=Z$ nucleus $^{66}$As \cite{Hasegawa05}. Our
analysis showed that there are essentially two different types of
isomer entering into the discussion. One of them is shape isomer
that occurs because prolate and oblate shapes can coexist at low
excitations along the $N=Z$ line. In fact, the prolate-oblate shape
coexistence is a well known phenomenon in the neighboring even-even
$N=Z$ nucleus $^{68}$Se \cite{Fischer}, where the ground state and
the first excited state have oblate and prolate deformation,
respectively \cite{Kaneko04,Sun04}. The first excited $0^{+}$ state,
which typically lies about several hundred keV above the ground
state, can decay to the ground state via an electric monopole (E0)
transition \cite{Bouchez}. The E0 transition is a very slow process,
and therefore, the first excited state becomes a shape isomer
\cite{Sun,Nature}. Thus, we may expect that the low-spin 7$^{+}$
state in $^{94}$Ag is a shape isomer in nature.

The $g_{9/2}$-shell nuclei were extensively studied in the early
years by the empirical shell model calculations
\cite{Gross,Serduke}. The very restricted model space
($1p_{1/2},0g_{9/2}$) was used as it allowed an empirical fit for
both residual interaction and single-particle energies. Herndl and
Brown \cite{HB} performed a detailed study of the $\beta$-decay
properties using the same model space. The interaction generally
yielded a good agreement with experimental data of the high-spin
spectroscopy. Later, the shell-model calculations with the $fpg$
model space comprising the ($1p_{3/2}$, $0f_{5/2}$, $1p_{1/2}$,
$0g_{9/2}$) shells together with a realistic interaction were
performed \cite{Schmidt}. The realistic effective interaction can in
principle be derived from the free nucleon-nucleon interaction, and
in fact, such microscopic interactions have been proposed for the
beginning of the shell \cite{Kuo,Jensen}. However, these
interactions failed to reproduce excitation spectra, binding
energies, and transitions if many valence nucleons are considered.
To overcome this defect, considerable effort has been put forward
with an empirical fit to experimental data \cite{Poves,Honma}.

On the other hand, realistic effective interactions are dominated
by pairing and multipole interactions with
the monopole term \cite{Dufour}. As documented in the literature, it
has been shown that the extended $P+QQ$ model works well for
a wide range of $N \approx Z$ nuclei \cite{Hasegawa01,Kaneko02}. This
model has demonstrated its capability of describing
the microscopic structure in different nuclei, as for instance, in
the $fp$-shell region \cite{Hasegawa01} and the $fpg$-shell region
\cite{Kaneko02}.

In this paper, we perform the spherical large-scale shell model
calculations in the $fpg$ model space for the upper $g_{9/2}$-shell
nuclei close to $N=Z$ line. The structure of the isomeric states in
the odd-odd $N=Z$ nucleus $^{94}$Ag is investigated in detail. In
particular, it is very interesting to study the deformation property
of the high-spin $21^{+}$ isomer. Our analysis shows that there are
essentially two different types of isomer entering into the
discussion. Due to the fact that prolate and oblate shapes can
coexist at the low excitation region, a shape isomer is suggested
for the low-spin $7^{+}$ state in $^{94}$Ag. In an agreement with
the previous conclusion, the high-spin $21^{+}$ isomer is
interpreted as a spin-gap isomer. The spherical large-scale shell
model calculations in the $gds$ model space comprising the
($0g_{9/2}$, $1d_{5/2}$, $0g_{7/2}$, $2s_{1/2}$) shells are carried
out to study the role of core excitations in the shell-model
structure associated with a 21$^{+}$-19$^{+}$ level inversion in
$^{94}$Ag.

The paper is arranged as follows. In Sec.~\ref{sec2}, we outline our
model. In Section~\ref{sec3}, the shell model results are presented
for several upper $g_{9/2}$-shell nuclei. In Section~\ref{sec4}, we
perform the numerical calculations and discuss the results for
$^{94}$Ag. Finally, conclusions are drawn in Section~\ref{sec5}.

\section{The model}\label{sec2}

We start with the following form of Hamiltonian, which consists of
pairing and multipole terms with the monopole interaction
\begin{eqnarray}
 H & = & H_{\rm sp} + H_{P_0} + H_{P_2} + H_{QQ} + H_{OO}
       + H^{T=0}_{\pi \nu} + H_{\rm mc}  \nonumber \\
   & = & \sum_{\alpha} \varepsilon_a c_\alpha^\dag c_\alpha
    -  \sum_{J=0,2} \frac{1}{2} g_J \sum_{M\kappa} P^\dag_{JM1\kappa} P_{JM1\kappa}  \nonumber \\
   & - & \frac{1}{2} \chi_2/b^{4} \sum_M :Q^\dag_{2M} Q_{2M}:
         - \frac{1}{2} \chi_3/b^{6} \sum_M :O^\dag_{3M} O_{3M}: \nonumber \\
   & - & k^0 \sum_{a \leq b} \sum_{JM} A^\dagger_{JM00}(ab) A_{JM00}(ab)
            \nonumber \\
   & + & \sum_{a \leq b} \sum_{T} k_{\rm mc}^T(ab) \sum_{JMK}
               A^\dagger_{JMTK}(ab) A_{JMTK}(ab),
          \label{eq:0}
\end{eqnarray}
where $b$ in the third and fourth terms is the length parameter of
harmonic oscillator. We take the $J=0$ and $J=2$ forces in the
pairing channel, and the quadrupole-quadrupole ($QQ$) and
octupole-octupole ($OO$) forces in the particle-hole channel
\cite{Hasegawa01,Kaneko02}. The monopole interaction is divided into
two parts, namely the average $T=0$ monopole field $H^{T=0}_{\pi
\nu}$ and the monopole correction term $H_{\rm mc}$. The Hamiltonian
(\ref{eq:0}) is isospin invariant, and is diagonalized in a chosen
model space based on a spherical basis \cite{Mizusaki}. In the
present work, we first employ the $fpg$ model space. This shell
model has proven to be rather successful in describing energy levels
and electromagnetic transitions. For the upper $g_{9/2}$-shell
nuclei, we employ the single-particle energies $\varepsilon_{p3/2} =
0.00$, $\varepsilon_{f5/2} = 0.77$,$\varepsilon_{p1/2} = 1.11$, and
$\varepsilon_{g9/2} = 3.70$ (all in MeV). We adopt the following
interaction strengths for the pairing and multipole forces
\begin{eqnarray}
 & {} &  g_0 = 24.0/A, \quad g_2 = 225.3/A^{5/3},            \nonumber \\
 & {} &  \chi_2 = 480.0/A^{5/3},
         \chi_3 = 368.6/A^2 (\mbox{ in MeV}), \label{eq:1}
\end{eqnarray}
and for the monopole terms
\begin{eqnarray}
  & {} & k_{\rm mc}^{T=0}(a,g_{9/2}) = -0.20 \mbox{ MeV},
 \quad a=p_{3/2}, f_{5/2}, p_{1/2},  \nonumber \\
 & {} & k_{\rm mc}^{T=1}(p_{3/2},f_{5/2}) = -0.3,
   \quad k_{\rm mc}^{T=1}(p_{3/2},p_{1/2}) = -0.3, \nonumber \\
 & {} & k_{\rm mc}^{T=1}(f_{5/2},p_{1/2}) = -0.4,
   \quad k_{\rm mc}^{T=1}(g_{9/2},g_{9/2}) = -0.1, \nonumber \\
 & {} & k_{\rm mc}^{T=0}(g_{9/2},g_{9/2}) = -0.75   \quad (\mbox{ in MeV}).
\label{eq:2}
\end{eqnarray}
We employ all the monopole terms used in our previous paper
\cite{Hasegawa051}, with additional terms $k_{\rm
mc}^{T=0}(a,g_{9/2})$ and $k_{\rm mc}^{T=0}(g_{9/2},g_{9/2})$
\cite{Hasegawa05}. These terms were found necessary for obtaining
correct positions of the $9_1^+$ and higher spin states relative to
the $7_1^+$ state in $^{66}$As. We have increased the strength of
the monopole term $k_{\rm mc}^{T=0}(g_{9/2},g_{9/2})$, which gives
effects of lowering the $g_{9/2}$ orbital and the negative parity
states. The $T=0$ monopole field $H^{T=0}_{\pi \nu}$ affects
significantly the relative energy between the $T=1$ and $T=0$ states
in odd-odd $N=Z$ nuclei \cite{Hasegawa04}. We have determined the
strength $k^0$ to be $64/A$ so as to reproduce the $7^{+}$ state at
the excitation energy 0.66 MeV in $^{94}$Ag. For calculations of
spectroscopic $Q$-moments and B(E2) values, we use the standard
effective charge $e_\pi=1.5e$ for protons and $e_\nu=0.5e$ for
neutrons.

\section{Calculation of upper $g_{9/2}$-shell nuclei in the
$fpg$ model space}
\label{sec3}

Before studying the odd-odd $N=Z$ nucleus $^{94}$Ag, we first
perform large-scale shell model calculations for several neighboring
$g_{9/2}$-shell nuclei to validate our model. The calculations are
carried out in the $fpg$ model space. The nuclei studied in this
section are the heaviest $N \approx Z$ nuclei for which energy
levels could be obtained experimentally. The structure information
of these nuclei is useful for the reaction network calculations of
the rp process.

\begin{figure}[t]
\includegraphics[totalheight=9.5cm]{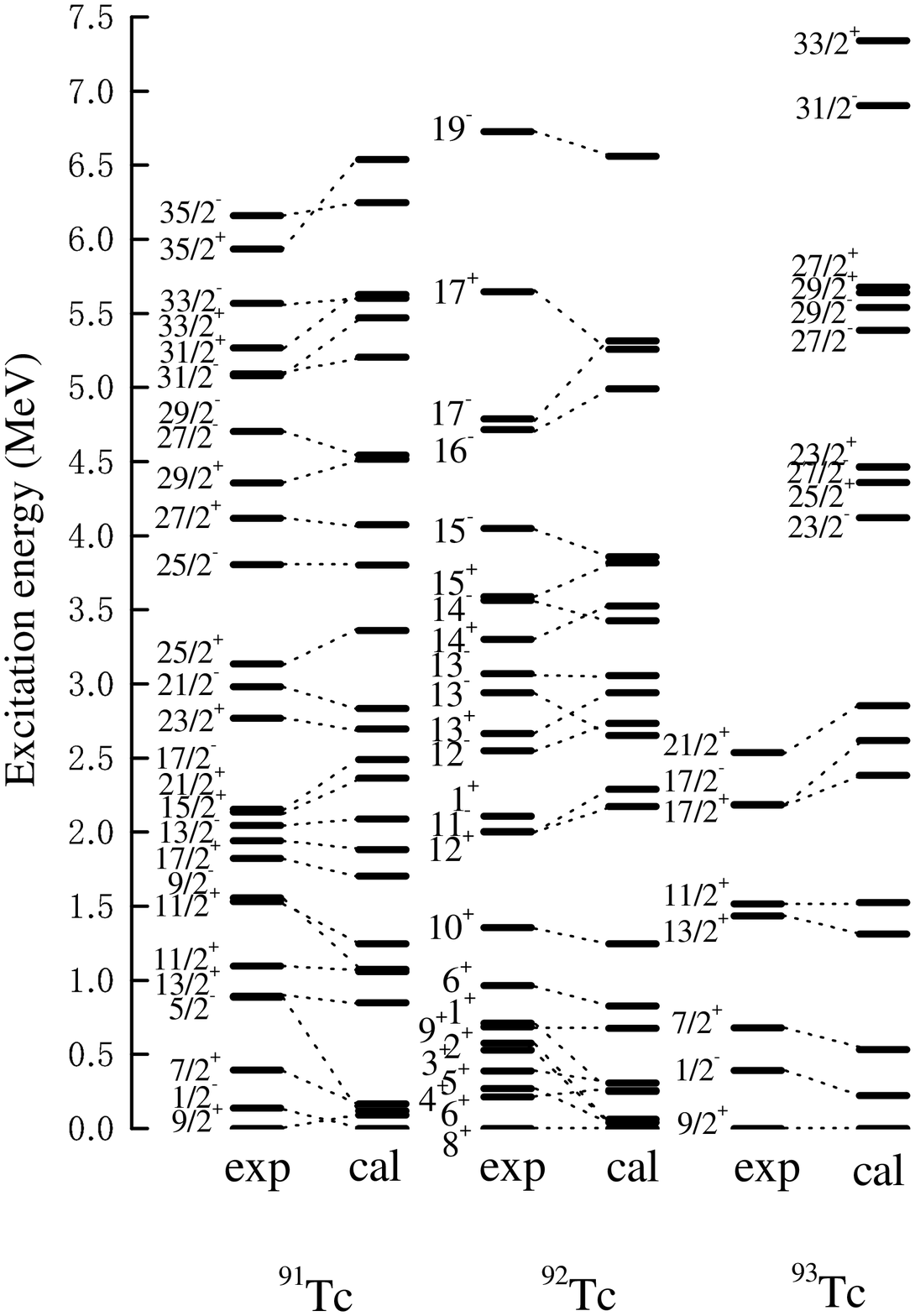}
  \caption{Experimental and calculated energy levels of the odd-even
nuclei $^{91}$Tc, $^{92}$Tc, and $^{93}$Tc.}
  \label{fig1}
\end{figure}

\begin{figure}[t]
\includegraphics[totalheight=9.5cm]{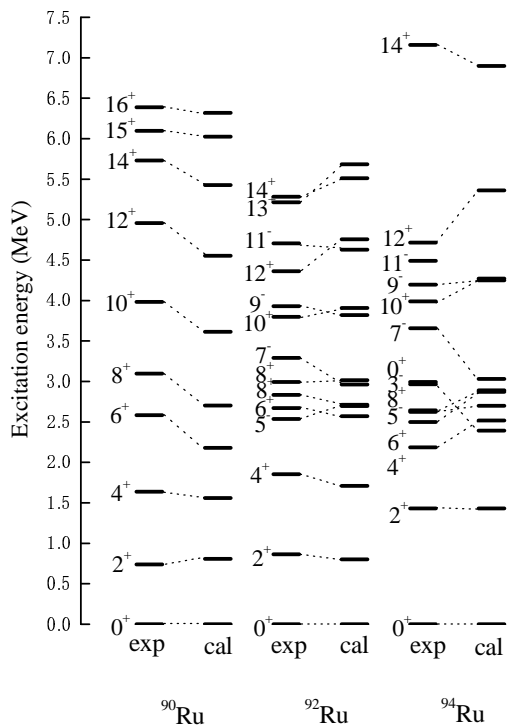}
  \caption{Experimental and calculated energy levels of the even-even
nuclei $^{90}$Ru, $^{92}$Ru, and $^{94}$Ru.}
  \label{fig2}
\end{figure}

\begin{figure}[t]
\includegraphics[totalheight=9.5cm]{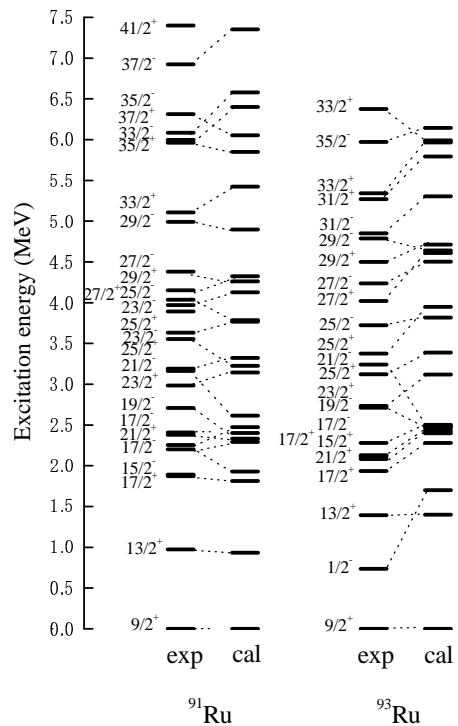}
  \caption{Experimental and calculated energy levels of the even-odd
nuclei $^{91}$Ru, and $^{93}$Ru.}
  \label{fig3}
\end{figure}

Figure \ref{fig1} shows the energy levels for the Tc isotopes. As
one can see, the calculated results reproduce satisfactorily the
experimental levels for both even and odd parity observed in the
odd-even nuclei $^{91,93}$Tc, and in the odd-odd nucleus $^{92}$Tc.

Comparison of the calculated energy levels with experimental data
for the Ru isotopes are shown in Fig.~\ref{fig2} for even-even nuclei, and
in Fig.~\ref{fig3} for odd-mass nuclei. Again, a good agreement with
experimental data is achieved. In Ref. \cite{Hasegawa04}, some
$N\approx Z$ Ru isotopes were investigated by using a slightly
different parameter set. Those calculations indicated an enhancement
of quadrupole correlations in the $N=Z$ nucleus $^{88}$Ru; however
the amount of the enhancement was not enough when compared the
calculated moments of inertia with the experimental values. It was
suggested that the $1d_{5/2}$ orbital in the next major shell
contributes very much to the quadrupole correlations. In fact, the
moment of inertia can be easily reproduced with the truncated model
space ($1p_{1/2}$,$0g_{9/2}$,$1d_{5/2}$) that includes the
$1d_{5/2}$ orbital. As we shall see in the next section,
contribution from the $1d_{5/2}$ orbital plays an important role
also in $^{94}$Ag.

\begin{figure}[t]
\includegraphics[totalheight=9.5cm]{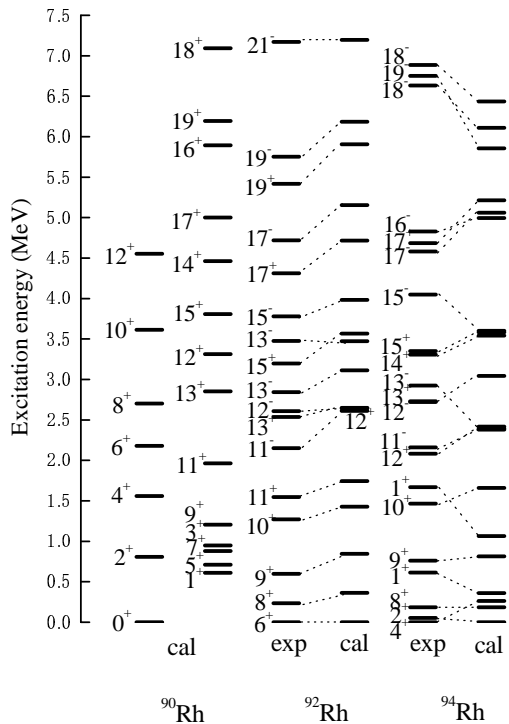}
  \caption{Experimental and calculated energy levels of the odd-odd
nuclei $^{90}$Rh, $^{92}$Rh, and $^{94}$Rh.}
  \label{fig4}
\end{figure}
\begin{figure}[t]
\includegraphics[totalheight=9.5cm]{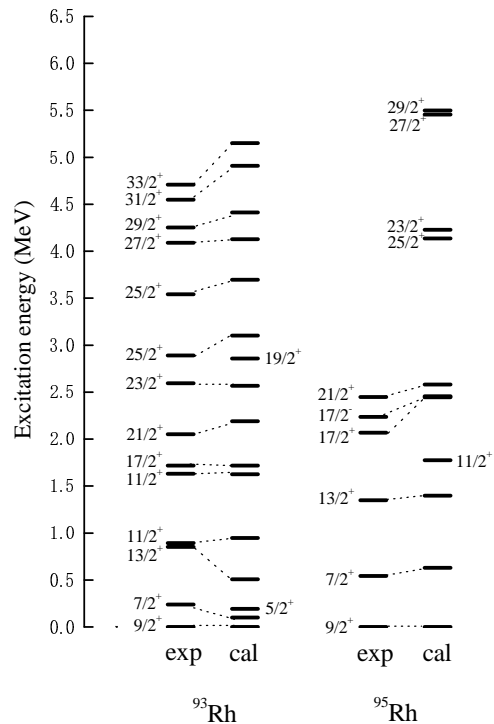}
  \caption{Experimental and calculated energy levels of the odd-even
nuclei $^{93}$Rh and $^{95}$Rh.}
  \label{fig5}
\end{figure}

Fig.~\ref{fig4} shows the calculated energy levels for the odd-odd
Rh isotopes $^{90,92,94}$Rh, and Fig.~\ref{fig5} for the odd-even
$^{93,95}$Rh. The results are compared with experimental data if
available. While the experimental levels of $^{92,93}$Rh are
correctly reproduced and the $^{95}$Rh data are too sparse to allow
a definite conclusion, our theoretical results for $^{94}$Rh are not
in good agreement with data. This disagreement might indicate the
importance of core excitation across the neutron shell-closure $N$ =
50 for this isotope, which is not taken into account in the present
model space. Currently, there is no experimental information for the
odd-odd $N=Z$ nucleus $^{90}$Rh. Our calculation thus serves as a
prediction, for the ground-state band with isospin $T=1$ and a side
band with $T=0$. It is interesting to mention that from our
calculation, the spectroscopic quadrupole moments are negative in
the ground-state band, while those in the side band are positive,
except for the $7^{+}$ state at an excitation about 1 MeV.
Therefore, we suggest that both the $T=0$ bandhead $1^{+}$ state and
the low-lying $7^{+}$ state may be considered as a candidate of
shape isomer.

\begin{figure}[t]
\includegraphics[totalheight=9.5cm]{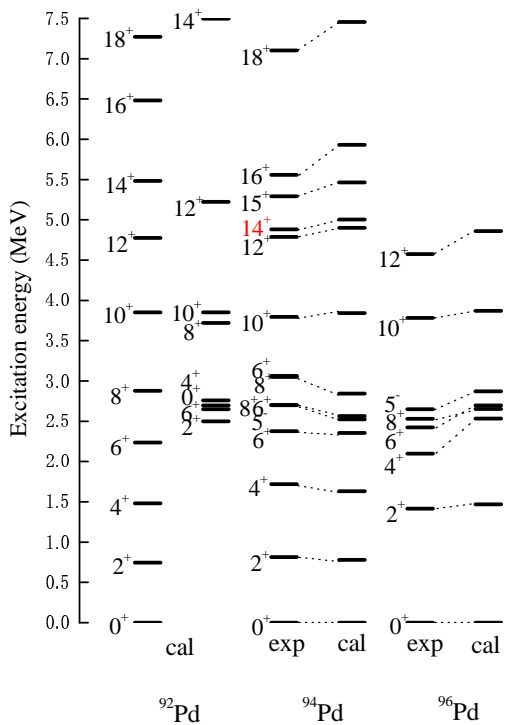}
  \caption{(Color online) Experimental and calculated energy levels
with $T=1$ of the even-even nuclei $^{92}$Pd, $^{94}$Pd, and $^{96}$Pd.}
  \label{fig6}
\end{figure}
\begin{figure}[t]
\includegraphics[totalheight=9.5cm]{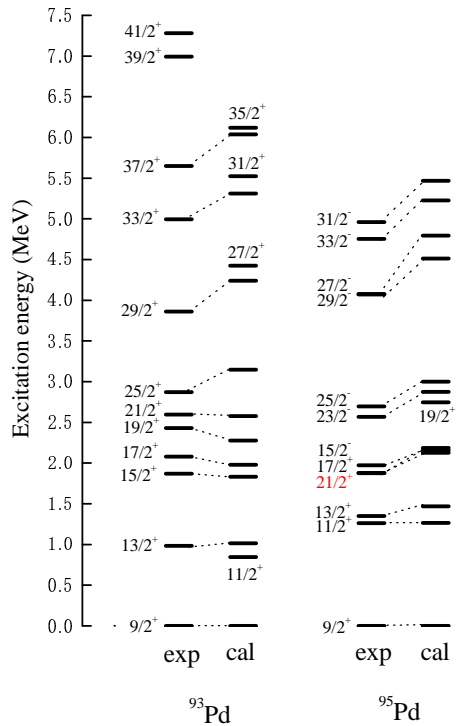}
  \caption{(Color online) Experimental and calculated energy levels
of the even-odd nuclei $^{93}$Pd and $^{95}$Pd.}
  \label{fig7}
\end{figure}

In Figs.~\ref{fig6} and \ref{fig7}, theoretical energy levels of the
Pd isotopes are compared with experiment. According to Ref.
\cite{Schatz1}, $^{92,93}$Pd are waiting point nuclei and their
structure information is important for the rp process
nucleosynthesis. As one can see from Figs.~\ref{fig6} and
\ref{fig7}, all the known energy levels for both even-even and
even-odd isotopes are well reproduced. It has been confirmed
experimentally that the $14^{+}$ state in $^{94}$Pd is an isomer
\cite{Marginean}. Using the standard effective charges 1.5$e$ for
protons and 0.5$e$ for neutrons, we have obtained $B(E2;14^{+}
\rightarrow 12^{+})$ = 39.6 $e^{2}fm^{4}$ for the isomeric $14^{+}$
state. This value is very close to the experimental estimate $B(E2)$
= 44(6) $e^{2}fm^{4}$ or $B(E2)$ = 39(9) $e^{2}fm^{4}$. In
Fig.~\ref{fig7}, our calculation predicts a low-lying ${11/2}^+$
level in $^{93}$Pd which has not been seen experimentally. The
theoretical level scheme for $^{95}$Pd correctly gives a
${21/2}^{+}$ isomeric state that is de-excited by an $E4$ decay. No
experimental information is currently available for the even-even
$N=Z$ nucleus $^{92}$Pd. The level scheme for $^{92}$Pd is thus our
prediction, where the calculated ground-state band and the side band
are suggested to have isospin $T=1$ and $T=0$, respectively.

\begin{figure}[t]
\includegraphics[totalheight=9.5cm]{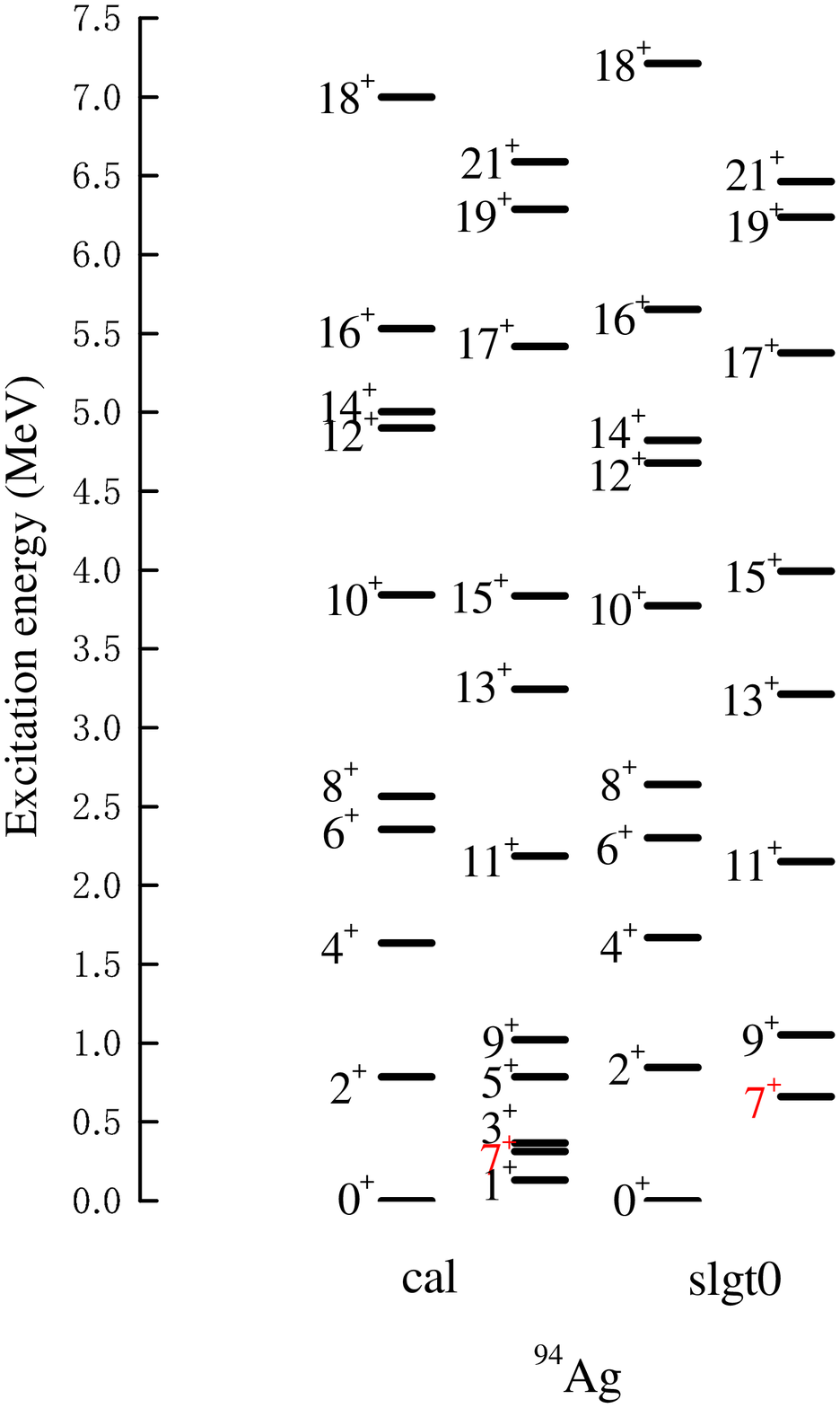}
  \caption{(Color online) Energy levels of the odd-odd
$N=Z$ nucleus $^{94}$Ag for the extended $P+QQ$ force in the $fpg$-shell
and the slg0 interaction in the restricted ($p_{1/2}$, $g_{9/2}$) model space. }
  \label{fig8}
\end{figure}

Figure \ref{fig8} shows the theoretical level scheme of the odd-odd
$N=Z$ nucleus $^{94}$Ag calculated within the present model space,
together with that obtained with the slgt0 effective interaction for
the (1$p_{1/2}$, $0g_{9/2}$) model space. The experimentally known
isomers in this nucleus include the low-spin $7^+$ and the high-spin
$21^{+}$ one. The question why these states become isomeric has, in
our opinion, not been thoroughly addressed. The excitation energy,
spin, and parity of the $7^+$ isomer have not been determined
experimentally. In the present calculation, we obtain a $7^{+}$
state around the excitation energy 0.66 MeV, which only has the
$0^+$ ground state and the $1^+$ state below it. Therefore, decay
out from the $7^{+}$ state can be strongly hindered by the selection
rule. This already creates a favorable condition for the $7^+$ state
to be isomeric. In Table I, we further analyze the structure of the
isomeric state by showing expectation values of nucleon number in
four orbitals as well as calculated spectroscopic quadrupole
moments. It is interesting to note that the results indicate a very
different structure of the $7^{+}$ state from the neighboring states
of $1^{+}$, $3^{+}$ and $5^{+}$. Opposite to all those neighboring
states, the $7^{+}$ state takes a quadrupole moment with positive
value. Hence the shape of the $7^{+}$ state is predicted to be
oblate, in contrast to the prolate shape for other neighboring
states. In this sense, the $7^{+}$ state can be considered to be a
shape isomer.

\begin{table}[b]
\caption{Expectation values of proton or neutron numbers occupied
         in the four orbitals, calculated for the low-lying $T=0$ states
         of $I^{\pi}$ in the $fpg$-shell model space.
         Calculated spectroscopic $Q$-moments (in $e$ fm$^2$) are
         also tabulated.}
\begin{tabular}{cccccccccc}   \hline\hline
$^{94}$Ag & \hspace{0.5cm}$p_{3/2}$\hspace{0.5cm} & \hspace{0.5cm}$f_{5/2}$\hspace{0.5cm}
 & \hspace{0.5cm}$p_{1/2}$\hspace{0.5cm} & \hspace{0.5cm}$g_{9/2}$\hspace{0.5cm}
          & \hspace{0.5cm}$Q$\hspace{0.5cm}  \\ \hline
 $1_1^+$ & 3.93 & 5.97 & 1.99 & 7.11 &  -16.6 \\
 $3_1^+$ & 3.93 & 5.97 & 2.00 & 7.10 &  -29.6 \\
 $5_1^+$ & 3.95 & 5.97 & 1.99 & 7.09 &  -26.7 \\
 $7_1^+$ & 3.95 & 5.96 & 2.00 & 7.08 &   61.2 \\
 $9_1^+$ & 3.98 & 5.95 & 1.99 & 7.08 &   23.8 \\
 $11_1^+$ & 3.56 & 5.92 & 1.98 & 7.66 &   44.5 \\
 $13_1^+$ & 3.75 & 5.77 & 1.91 & 7.57 &   55.0 \\
 $15_1^+$ & 3.98 & 5.44 & 1.98 & 7.60 &   55.2 \\
 $17_1^+$ & 3.83 & 5.68 & 1.96 & 7.52 &   43.6 \\
 $19_1^+$ & 3.82 & 5.72 & 1.95 & 7.51 &   48.3 \\
 $21_1^+$ & 4.00 & 5.50 & 2.00 & 7.50 &   57.7 \\ \hline\hline
\end{tabular}
\label{table1}
\end{table}

\begin{figure}[t]
\includegraphics[totalheight=9.5cm]{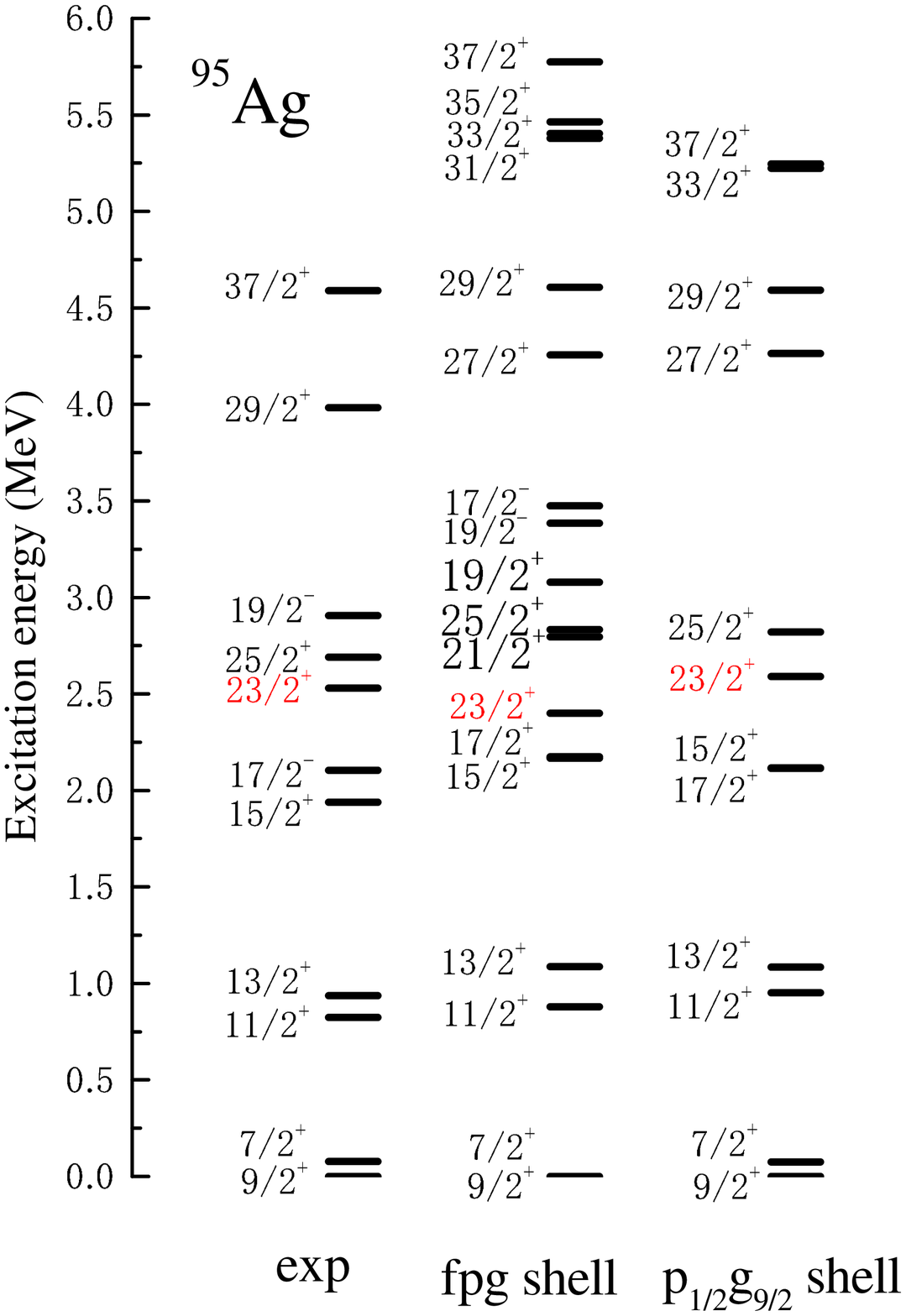}
  \caption{(Color online) Experimental and calculated energy levels
of $^{95}$Ag in the $fpg$-shell and the ($p_{1/2}$, $g_{9/2}$) model space.}
  \label{fig9}
\end{figure}

Calculated energy levels for the odd-mass nucleus $^{95}$Ag are
shown in Fig. \ref{fig9}. The calculation has well reproduced the
known experimental levels in the low-spin part while the theoretical
levels of the high spin states are too high as compared to data. The
calculation produces a level inversion in which the $23/2^{+}$ state
lies below the $21/2^{+}$ state. Thus, decay of the $23/2^{+}$ level
has to go with an E4/M3 transition. This may suggest a possible
isomerism of the $23/2^{+}$ state. However, our calculation cannot
explain the experimental finding that the $23/2^{+}$ state decays to
the $17/2^{-}$ one through an E3 transition.

\begin{figure}[t]
\includegraphics[totalheight=9.5cm]{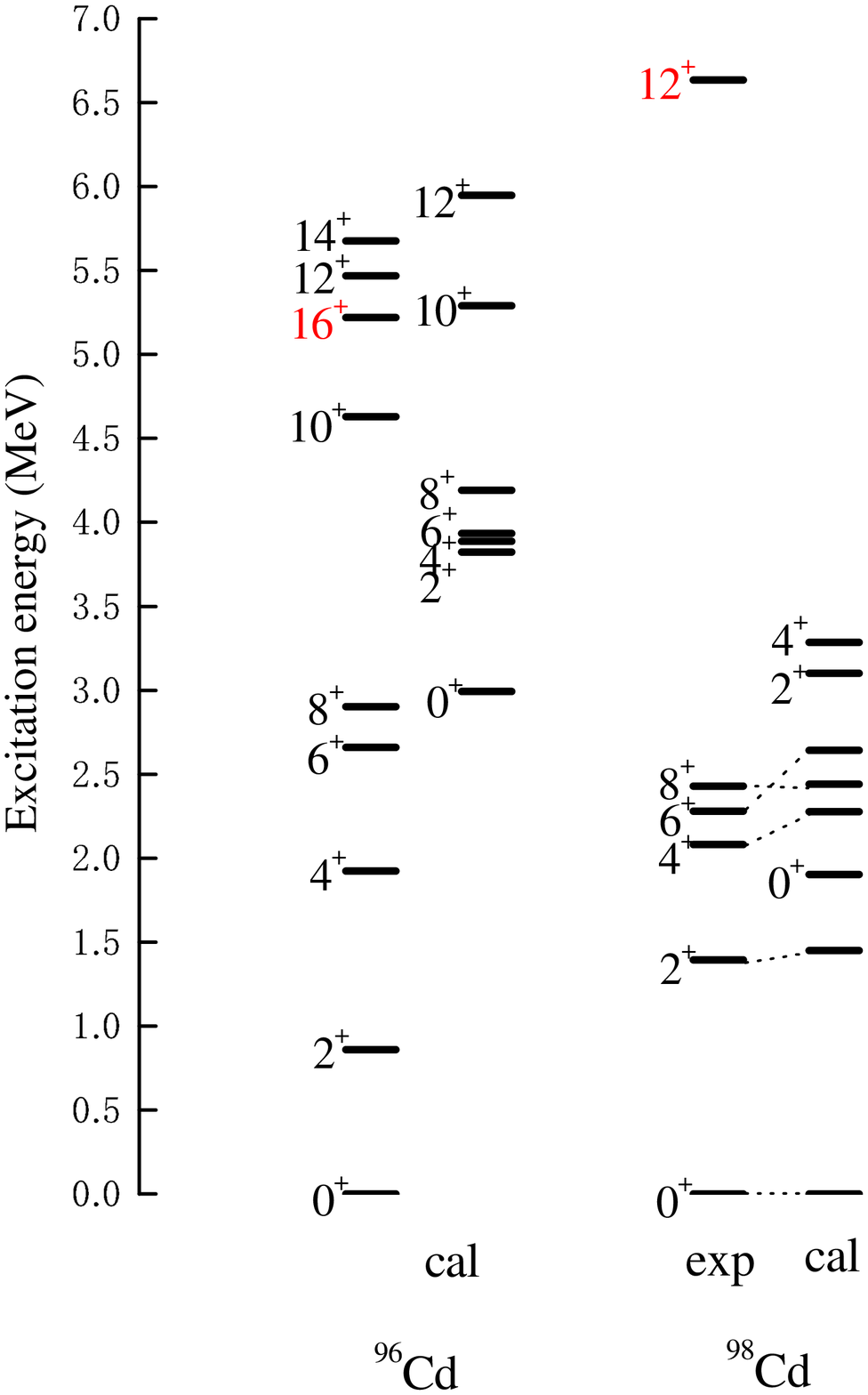}
  \caption{(Color online) Experimental and calculated energy levels of
the even-even nuclei $^{96}$Cd and $^{98}$Cd.}
  \label{fig10}
\end{figure}

Finally, theoretical energy levels of the Cd isotopes are shown in
Fig.~\ref{fig10}. The experimentally known energy levels for
$^{98}$Cd are well reproduced.
In recent experiment \cite{Blazhev}, the $12^{+}$ isomer was identified
at 6.635 MeV in $^{98}$Cd. In the present calculations, however,
the $12^{+}$ state is out of the $fpg$ shell configurations and
the upper shells such as $1d_{5/2}$ orbital are needed.
For the even-even $N=Z$ nucleus
$^{96}$Cd, which has not been investigated experimentally, our model
yields the ground and side bands as shown in Fig.~\ref{fig10}. It
is worthwhile to mention that the present calculation predicts a
spin-gap isomer with spin-parity $16^{+}$ at about 5.2 MeV, in
consistent with the shell model result obtained in the restricted
($1p_{1/2}$,$0g_{9/2}$) model space. $^{96}$Cd is known as a waiting
point nucleus \cite{Schatz1}.

\section{Calculation with the $gds$ model space
for $^{94}$Ag}\label{sec4}

We have demonstrated the power of our $fpg$ shell model in a
systematical description of proton-rich upper $g_{9/2}$-shell
nuclei. For $^{94}$Ag, the model has explained the structure of the
low-lying isomer $7^{+}$. The calculation with this ($1p_{3/2}$,
$0f_{5/2}$, $1p_{1/2}$, $0g_{9/2}$) model space, however, could not
reproduce the high-spin isomer $21^{+}$ in $^{94}$Ag. It gives a
quite large electric quadrupole transition value,
$B(E2;21^{+}\rightarrow 19^{+})$ = 58.9 $e^{2}fm^{4}$, which
corresponds to a fast decay to the lower state $19^{+}$. This result
is similar to that obtained in the shell model calculation with the
restricted ($1p_{1/2}$, $0g_{9/2}$) model space \cite{Serduke}. On
the other hand, it has recently been reported by Plettner {\it et
al.} \cite{Plettner} that the shell model calculation in the
($0g_{9/2}$, $1d_{5/2}$, $0g_{7/2}$, $1d_{3/2}$, $2s_{1/2}$) model
space can predict the isomerism of the $21^{+}$ state due to a
$21^{+}$-$19^{+}$ level inversion. The work of Plettner {\it et al.}
thus suggests an important contribution from the upper orbitals
above $0g_{9/2}$.

In a previous paper \cite{Hasegawa04} that discusses the structure
of the even-even $N=Z$ nucleus $^{88}$Ru, some of us suggested that
the $1d_{5/2}$ orbit above the $fpg$-shell plays a significant role
in the collectivity of that nucleus. The $1d_{5/2}$ orbit is
expected to contribute to the quadrupole correlation because it
couples strongly with the $0g_{9/2}$ orbit through the $Q$ matrix
element $\langle 0g_{9/2} || Q || 1d_{5/2} \rangle$ with $\Delta l=
\Delta j =2$. Simply adding the $1d_{5/2}$ orbit to the current
model space $(1p_{3/2},0f_{5/2},1p_{1/2},0g_{9/2})$, unfortunately,
makes the configuration space too large. In order to include the
$1d_{5/2}$ orbit, we have to truncate out some lower orbits that may
not be very important for the present discussion. Table I has shown
that for the upper $g_{9/2}$-shell nuclei, the contribution from the
$(1p_{3/2},0f_{5/2},1p_{1/2})$ shells is not so large as compared to
the $0g_{9/2}$ orbit. Accordingly, we shall examine the contribution
of the $1d_{5/2}$ orbit within the truncated space
$(0g_{9/2},1d_{5/2},0g_{7/2},2s_{1/2})$, hereafter called the $gds$
model space. This model space, without $(1p_{3/2},0f_{5/2},1p_{1/2})$,
is expected to work properly as it can explain the main features of the
upper $g_{9/2}$-shell nuclei \cite{Zuker,Caurier}.

Let us now perform large-scale shell model calculations within the
$gds$ model space, which allows excitations across the $N=Z=50$
shell gap. For the single-particle energies, we employ the predicted
ones from the global fit of all available single-particle and
single-hole energies \cite{Nowacki,Grawe} of Duflo and Zuker
\cite{Duflo}, i.e., $\varepsilon_{g9/2} = 0.0$, $\varepsilon_{d5/2}
= 2.54$, $\varepsilon_{g7/2} = 4.95$, and $\varepsilon_{s1/2} =
3.34$ in MeV. For the interaction strengths in the Hamiltonian, the
following constants are adopted as
\begin{eqnarray}
 & {} &  g_0 = 24.0/A, \quad g_2 = 225.3/A^{5/3},  \nonumber \\
 & {} &  \chi_2 = 240.0/A^{5/3} \mbox{ in MeV}. \label{eq:3}
\end{eqnarray}
The octupole-octupole force and the monopole correction terms are
neglected because without these terms, the essential feature of level
sequencies does not change.
The inclusion of $1d_{5/2}$ orbit allows us to use a weaker
quadrupole force, and therefore, $\chi_2$ in (\ref{eq:3}) is smaller
than that in Eq. (\ref{eq:1}). It is interesting to note that these
force strengths are close to those employed in our previous paper
\cite{Hasegawa05}. The $T=0$ monopole field $H^{T=0}_{\pi \nu}$
influences strongly the relative energy between the $T=1$ and $T=0$
states in odd-odd $N=Z$ nuclei \cite{Hasegawa04}. The strength $k^0
=64/A$ is chosen so as to reproduce the excitation energy 0.66 MeV
of the low-lying $7^{+}$ state. Again, we take the standard
effective charges $e_\pi=1.5e$ for protons and $e_\nu=0.5e$ for
neutrons.

\begin{figure}[t]
\includegraphics[totalheight=8.5cm]{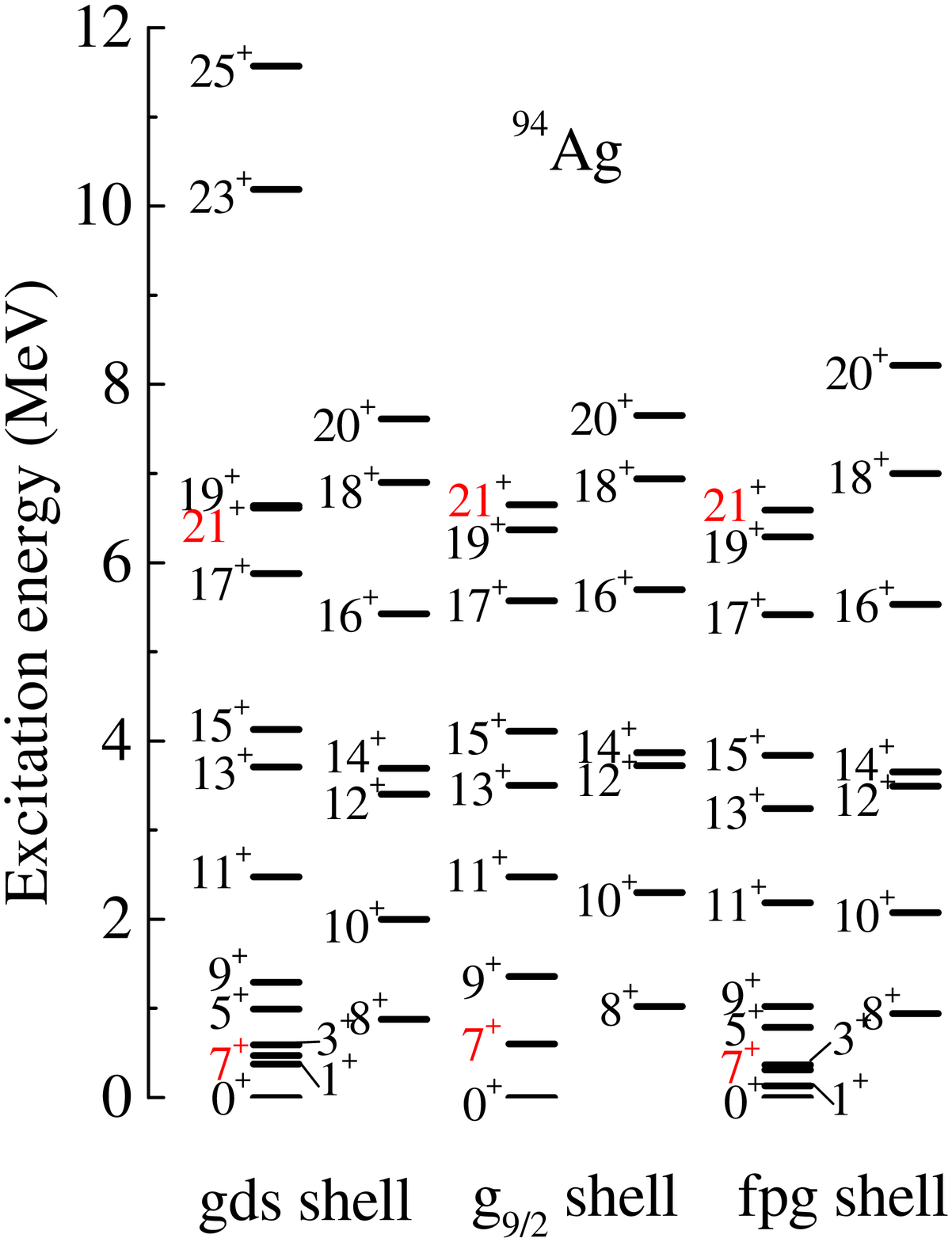}
  \caption{(Color online) Calculated energy levels with $T=0$ of the odd-odd
$N=Z$ nucleus $^{94}$Ag with the $gds$-shell, $g_{9/2}$-shell, and
$fpg$-shell model spaces.}
  \label{fig11}
\end{figure}

Calculations are performed by allowing up to 3p-3h excitations
across the $N=Z=50$ shell gap, which are sufficient for the present
discussion. Energy levels of $^{94}$Ag obtained from this
calculation are shown in Fig. \ref{fig11}. Results from different
model spaces are also given for comparison. As one can see, the
three results are quite similar, and they all reproduce well the few
known experimental energy levels. The closely-lying $12^{+}$ and
$14^{+}$ states with $T=1$ may well correspond to those discussed in
$^{94}$Pd (see Fig. 6). However, one clear difference among the
three results is that the $gds$-shell model calculation yields a
$21^{+}$-$19^{+}$ level inversion, with the $19^{+}$ level lying
25.7 keV above the $21^{+}$ level. The calculated E4 transition
probability $B(E4;21^{+}\rightarrow 17^{+})$ is very small. We can
thus explain the isomeric nature of the $21^{+}$ state within the
$gds$ shell model, in accordance with the result of Plettner {\it et
al.} \cite{Plettner} in the
($0g_{9/2}$,$1d_{5/2}$,$0g_{7/2}$,$2s_{1/2}$,$1d_{3/2}$) model
space. We mention that there is an interesting precedent, the $12^+$
isomer in $^{52}$Fe. As already noted in Ref. \cite{Poves}, this
phenomenon is due to jumps from the $f_{7/2}$ orbital to the above
$fp$-shell. Similarly, we can consider that the $21^+$ isomerism is
due to jumps from the $g_{9/2}$ orbital to the above $sdg$-shell.

\begin{figure}[t]
\includegraphics[totalheight=6.5cm]{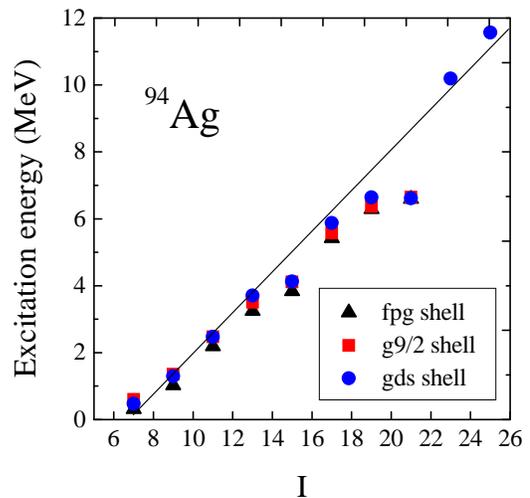}
  \caption{(Color online) Excitation energies of the $T=0$ states in
$^{94}$Ag from the $gds$-shell, $g_{9/2}$-shell, and $fpg$-shell model
calculations. The diagonal line is a linear fit.}
  \label{fig12}
\end{figure}

Excitation energies of the $T=0$ states are plotted as functions of
total spin $I$ in Fig.~\ref{fig12}. The figure displays an
approximate linear dependence of excitation energy on the total spin
$I$, with a clear deviation of the $21^{+}$ state from the straight
line. Thus, our shell model calculations demonstrate the importance
of excitations across the $^{100}$Sn closed shell, which causes a
$21^{+}$-$19^{+}$ level inversion, and hence the $21^+$ isomerism.

\begin{table}[b]
\caption{Expectation values of proton or neutron numbers occupied
         in the four orbitals, calculated for the low-lying $T=0$
         states of $I^{\pi}$ in the $gds$-shell model space.
         Calculated spectroscopic $Q$-moments (in $e$ fm$^2$) are
         also tabulated.}
\begin{tabular}{cccccccccc}   \hline\hline
$^{94}$Ag & \hspace{0.5cm}$g_{9/2}$\hspace{0.5cm} & \hspace{0.5cm}$d_{5/2}$\hspace{0.5cm}
 & \hspace{0.5cm}$g_{7/2}$\hspace{0.5cm} & \hspace{0.5cm}$s_{1/2}$\hspace{0.5cm}
          & \hspace{0.5cm}$Q$\hspace{0.5cm}  \\ \hline
 $1_1^+$ & 6.50 & 0.43 & 0.04 & 0.03 &  -30.7 \\
 $3_1^+$ & 6.51 & 0.43 & 0.04 & 0.03 &  -52.0 \\
 $5_1^+$ & 6.52 & 0.41 & 0.04 & 0.03 &  -53.2 \\
 $7_1^+$ & 6.51 & 0.43 & 0.04 & 0.03 &  109.6 \\
 $9_1^+$ & 6.55 & 0.38 & 0.04 & 0.03 &   43.0 \\
 $11_1^+$ & 6.56 & 0.37 & 0.04 & 0.03 &   11.1 \\
 $13_1^+$ & 6.62 & 0.32 & 0.04 & 0.02 &    9.2 \\
 $15_1^+$ & 6.65 & 0.29 & 0.04 & 0.02 &   70.5 \\
 $17_1^+$ & 6.70 & 0.26 & 0.03 & 0.02 &   39.3 \\
 $19_1^+$ & 6.72 & 0.23 & 0.03 & 0.02 &   53.3 \\
 $21_1^+$ & 6.70 & 0.26 & 0.03 & 0.02 &   84.1 \\ \hline\hline
\end{tabular}
\label{table2}
\end{table}

\begin{figure}[t]
\includegraphics[totalheight=6.5cm]{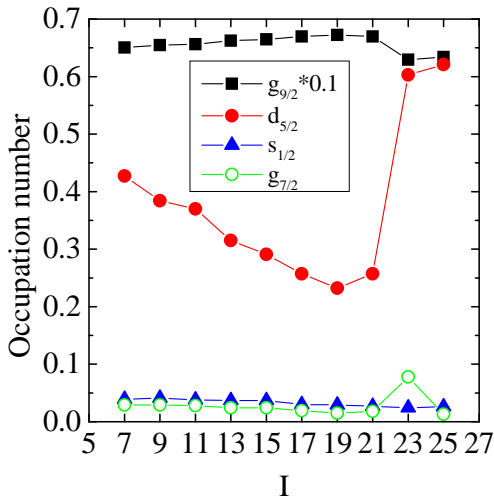}
  \caption{(Color online) Calculated occupation number as a function of
total spin in $^{94}$Ag for the $gds$-shell model calculations. The occutation
  numbers of $0g_{9/2}$ orbital are plotted as 0.1 times of their actual values.}
  \label{fig13}
\end{figure}
\begin{figure}[t]
\includegraphics[totalheight=6.5cm]{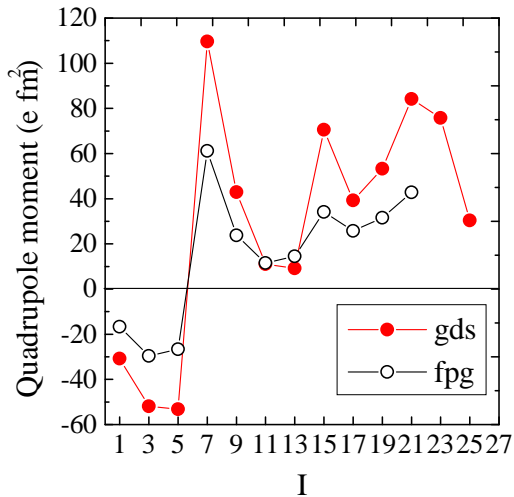}
  \caption{(Color online) Spectroscopic quarupole moments as a function
of total spin. }
  \label{fig14}
\end{figure}

\begin{table}[b]
\caption{$B(E2:J_i \rightarrow J_f)$ for the $T=0$ and $T=1$ states of $^{94}$Ag. }
\begin{tabular}{ccccccc}   \hline
  & \multicolumn{6}{c}{$B(E2:J_i \rightarrow J_f)$ in $e^2$fm$^4$} \\
  $J_i \rightarrow J_f$ & \hspace{0.3cm}$fpg$-shell\hspace{0.3cm}   &  \hspace{0.3cm}$gds$-shell\hspace{0.3cm}  \\ \hline\hline
   $T=0$            &     &      \\
  $3 \rightarrow 1$     & 474 & 669  \\
  $5 \rightarrow 3$     & 364 & 757  \\
  $7 \rightarrow 5$     & 0.0 &  0.26 \\
  $9 \rightarrow 7$     & 41  & 78  \\
  $11 \rightarrow 9$    & 139 & 298  \\
  $13 \rightarrow 11$   & 93  & 361  \\
  $15 \rightarrow 13$   & 63  & 88  \\
  $17 \rightarrow 15$   & 43  & 103  \\
  $19 \rightarrow 17$   & 103 & 130  \\
  $21 \rightarrow 19$   &  59 & 109  \\

  $T=1$             &     &     \\
  $2 \rightarrow 0$     & 193 & 497  \\
  $4 \rightarrow 2$     & 250 & 719  \\
  $6 \rightarrow 4$     & 247 & 718  \\
  $8 \rightarrow 6$     &  32 &  17  \\ \hline\hline
\end{tabular}
\label{table3}
\end{table}

With our shell model results, we now further examine the microscopic
structure of the isomeric $7^{+}$ and $21^{+}$ states in $^{94}$Ag.
In Table II, we present the expectation values of particle numbers
in four orbitals, calculated for the low-lying $T=0$ states in
$^{94}$Ag. Theoretical spectroscopic $Q$-moments (in $e$ fm$^2$) are
also included in the table. To be better visualized, these
occupation numbers are plotted as functions of total spin $I$ in
Fig. 13,
where for the $0g_{9/2}$ orbital, the plotted numbers are 0.1 times
of their actual values.
The results show that the most occupied orbit is
$0g_{9/2}$. The occupation number of the $1d_{5/2}$ orbit is 0.43
for the $7^{+}$ state. With increasing $I$, it decreases linearly,
but begins to rise at $21^{+}$. For the high-spin states $23^{+}$
and $25^{+}$, a drastic increase of $1d_{5/2}$ occupation is
observed. The $21^{+}$ state has the maximum aligned hole
configurations in the $fpg$ shell for neutrons and protons, which
coupled strongly with the $1d_{5/2}$ configurations. The extra
binding energy due to the large attractive {\it pn} interaction in
the $0g_{9/2}$ orbit lowers the $21^{+}$ state primarily. However,
this is not sufficient to make the $21^{+}$ state isomeric. It is
the mixing with the $1d_{5/2}$ configurations that eventually causes
the $21^{+}$-$19^{+}$ level inversion so that the $21^{+}$ state
becomes an isomer.

In Fig. \ref{fig14}, we plot the two calculations of spectroscopic
$Q$-moments for the $T=0$ excited states as a function of total spin
$I$, using the numbers listed in Tables I and II. Comparing the
$fpg$-shell and $gds$-shell model calculations, one sees a clear
enhancement in $Q$-moment at the $7^{+}$, $15^{+}$, and $21^{+}$
states due to the excitations across the $^{100}$Sn shell-closure.
We thus conclude that the inclusion of the $1d_{5/2}$ orbit enhances
significantly the collectivity of the isomeric $7^{+}$ and $21^{+}$
states. In the $gds$-shell calculation, the spectroscopic
$Q$-moments for $7^{+}$ and $21^{+}$ states are predicted to be
109.6 $e$ fm$^2$ and 84.1 $e$ fm$^2$, respectively.

The above conclusion is further supported by B(E2) calculations
shown in Table III. There, the B(E2) values are presented along the
two decay sequences with $T=0$ and $T=1$. What we can learn from the
table is the fact that the numbers in the column of the $gds$-shell
calculation are always larger than those of the $fpg$-shell results.
This indicates that the core excitations across the $N=Z=50$ shell
gap increase the B(E2) values, which is consistent with the
observation in Fig. 14. Moreover, the transition B(E2, $7^{+}\rightarrow
5^{+})$ is very small also in the $gds$-shell calculation, implying that
the core excitation does not wash out the isomeric nature of the
$7^+$ state as discussed in the $fpg$-shell calculation.

The analysis on one- and two-proton decay of the 21$^{+}$ isomer has
suggested a large deformation for this state. Mukha {\it et al.}
\cite{Mukha06} argued that the unexpectedly large probability for
the proton decay is attributed to a large prolate shape of the
parent nucleus. We note that shape is a widely-used concept but not
a direct shell model consequence. In order to discuss shapes using
the present shell model results, assumptions are needed. By assuming
an axial symmetric rotor for $^{94}$Ag, deformations for the
isomeric $7^{+}$ and $21^{+}$ states can be estimated to be $\beta =
-0.5$ and $\beta = -0.2$ from the spectroscopic $Q$-moments 109.6
$e$ fm$^2$ and 84.1 $e$ fm$^2$, respectively, if a $K=0$ is further
assumed for both states. While the assumption of $K=0$ may be
reasonable for the $7^{+}$ state, it leads to a conclusion that the
$21^{+}$ state corresponds to an oblate shape, in contradiction to
the conclusion of Ref. \cite{Mukha06}. However, a prolate shape of
$\beta = +0.2$ can be obtained if, for example, a $K=21$ is assumed
for the $21^{+}$ state.
The present calculations in the $gds$-shell do not show strong
deformation. As suggested by I. Mukha {\it et al.} \cite{Mukha05},
the $1h_{11/2}$ single-proton orbital may be important for a
strongly deformed shape of the $21^{+}$ isomer in $^{94}$Ag.
Independent of these assumptions, a
measurement of the laboratory quadrupole moments for the isomeric
states, which may be compared directly to our theoretical results
listed in Tables I and II, is desired.

\section{Conclusions}\label{sec5}

In this paper, we have investigated the microscopic structure of
isomers in $^{94}$Ag within a large-scale shell model framework. To
demonstrate that our model is reliable with predictive power, we
have systematically calculated other nuclei in the same mass region,
for all even-even, even-odd, odd-even, and odd-odd types of nuclei,
and compared the results with available data. Overall, a good
agreement has been obtained. For those $N=Z$ nuclei (in addition to
$^{94}$Ag, also $^{90}$Rh, $^{92}$Pd, and $^{96}$Cd) for which no
experimental information is available, the calculations may serve as
predictions for future experiment. In particular, the discussed
structure for those waiting point nuclei may be useful for the
reaction network calculations of the rp process nucleosynthesis.

We have shown that the isomerism of the $7^{+}$ and $21^{+}$ states
in $^{94}$Ag is attributed to an enhancement of collectivity that
causes large deformation. The isomeric $7^{+}$ state has been
suggested as a shape isomer because this isomerism is due to shape
difference between this and the lower-lying states. The $21^{+}$
isomer is produced by a different mechanism, namely, a level
inversion of the $19^{+}$ and $21^{+}$ states from their normal
order. We have found that this unusual level inversion is caused by
excitations across the $^{100}$Sn shell closure to the upper shells
including the $1d_{5/2}$ orbital. This conclusion is in an agreement
with the previous finding by Plettner {\it et al.} \cite{Plettner}.

It was discussed by Mukha {\it et al.} \cite{Mukha06} that the fine
structure of the observed proton decay would require a large
deformation for the $21^{+}$ isomeric state, and the unexpectedly
large probability for the observed two-proton decay could be
attributed to a strongly deformed prolate shape in $^{94}$Ag. The
present shell model calculation does not yield a large deformation
for the $21^{+}$ isomer, and it cannot provide a conclusive support
for the claim about a definite shape because in order to link the
shell model results to a geometric shape, one has to introduce
assumptions similar as those utilized in Ref. \cite{Mukha06}. 
The present result indicating no huge deformation of $^{94}$Ag isomer
could be reasonable because the ground states of nuclei in this region
are very weakly deformed. 
It remains an open question of whether a more extended shell model
space can obtain a large deformation. Our calculations predict that
the $21^{+}$ isomer has large positive value of the spectroscopic
$Q$-moment. To confirm this, the nuclear $Q$-moment should be
measured directly.

Y.S. is supported by the Chinese Major State Basic Research
Development Program through grant 2007CB815005, and by the the U. S.
National Science Foundation through grant PHY-0216783.



\end{document}